# Large-Area Transfer of 2D TMDCs Assisted by Water-soluble layer for Potential Device Applications

*Madan Sharma[1], Aditya Singh[1], Pallavi Aggarwal[1] and Rajendra Singh[1,2]\**

[1] Department of Physics, Indian Institute of Technology Delhi, Hauz Khas, New Delhi 110016, India

[2] Nanoscale Research Facility (NRF), Indian Institute of Technology Delhi, Hauz Khas, New Delhi 110016, India

## Abstract

Layer transfer offers enormous potential for the industrial implementation of 2D material technology platforms. However, the transfer method used must retain as-grown uniformity and cleanliness in the transferred films for the fabrication of 2D material-based. Additionally, the method used must be capable of large-area transfer to maintain wafer-scale fabrication standards. Here, a facile route to transfer centimeter-scale synthesized 2D TMDCs (3L $MoS_2$, 1L $WS_2$) onto various substrates such as sapphire, $SiO_2$/Si, and flexible substrates (mica, polyimide) has been developed using a water-soluble layer ($Na_2S/Na_2SO_4$) underneath the as-grown film. The developed transfer process represents a fast, clean, generic, and scalable technique to transfer 2D atomic layers. The key strategy used in this process includes the dissolution of $Na_2S/Na_2SO_4$ layer due to the penetration of NaOH solution between the growth substrate and hydrophobic 2D TMDC film. As a proof-of-concept device, a broadband photodetector has been fabricated onto transferred 3L $MoS_2$, which shows photoresponse behavior for a wide range of wavelength ranging from NIR to UV. The enhancement in photocurrent was found to be 100 times and 10 times to dark current in the UV and visible region, respectively. This work opens up the pathway towards flexible electronics and optoelectronics.

# 1. Introduction

Silicon-based electronic devices pose critical challenges in sub-10 nm regime [1]. Rapid miniaturization of electronic devices needs exploration of new semiconducting materials. In the recent decade, two-dimensional (2D) transition metal dichalcogenides (TMDCs) are seen as a new hope for the electronic market due to their exotic mechanical, electrical, thermal, and optical properties [2-4]. Indirect to direct bandgap cross-over of TMDCs from bulk to monolayer makes them suitable semiconducting materials for nanoscale electronic devices [5]. In the last few years, researchers are optimistic that TMDCs could be the potential replacement of silicon (Si) in front-end-of-line (FEOL) technologies due to spectacular improvement in their performance [6]. For example, the Hall mobility of 6L $MoS_2$ has been achieved up to 34,000 $cm^2$ $V^{-1}$ $s^{-1}$ at low temperature [7], and room temperature (RT) current on/off ratio of 1L $MoS_2$ can reach up to $\sim 10^8$ [8]. Initially, TMDCs-based devices have been fabricated onto exfoliated flakes or some micrometer-sized flakes to optimize the material quality and device performance. However, synthesis of large-area TMDCs is required to realize their commercial applications, so; considerable efforts have been given to grow the large-area 2D TMDCs.

Recently, *Yang et al.* have synthesized 6-inch large and uniform 1L $MoS_2$ by CVD on soda-lime glass at growth temperature 720 °C [9]. However, the growth requirements for large-area, high-quality materials, namely high growth temperatures (> 700 °C), choice of substrates, precursors used for growth, etc., limit the potential applications [10-12]. Exemplifying this, the growth process negatively affects the underlying substrate at high temperatures, which degrades the performance of fabricated devices [13, 14]. In contrast, the growth temperature can be reduced by choosing appropriate precursors that avoid unwanted doping and contamination on the growth substrate [15, 16]. Layer transfer of 2D TMDCs from the growth substrate to the application substrate is a viable approach to overcome these issues. Layer transfer is also needed for the fabrication of 2D/3D heterojunction-based devices [17-19], flexible and transparent devices [20, 21]. Due to good optical transparency, high strain limit, and the high surface-area-to-volume ratio [22, 23], 2D TMDCs are perfectly suited for use in flexible and wearable electronics and for Internet of Things (IoT) applications [16]. Significant efforts have been made to transfer 2D-material films without degrading the film quality, which is critical for the success of 2D materials [24-28]. Still, improvement and automation of layer transfer processes are needed for its industrial implementation.

In this work, we have developed a nondestructive, crack-free, and clean transfer technique based on the dissolution of $Na_2S/Na_2SO_4$ layer. Trilayer (3L) $MoS_2$ and monolayer (1L) $WS_2$ have been transferred onto arbitrary substrates using this layer transfer method. Furthermore, we have demonstrated the fabrication of photodetector onto transferred 3L $MoS_2$ to show the potential of this process in device applications. Photodetector shows significant photoresponse for a wide range of wavelengths ranging from NIR to UV. The photo-to-dark current ratio (PDCR) is 100 and 10 in the UV and visible region, respectively. Broadband photoresponse (NIR-Vis-UV) of trilayer $MoS_2$ has been observed in this work.

## 2. Experimental section

### 2.1. Synthesis of large-area trilayer $MoS_2$

Single zone APCVD system with a long quartz tube (45 cm) having 5 cm diameter was used to achieve the large-area synthesis of trilayer (3L) $MoS_2$ over $SiO_2$ (300 nm, thermally oxidized Si)/Si substrate. Before the synthesis, the tube was purged at 300 °C with 480 sccm argon (Ar) gas flow for 10 minutes to remove preoccupied precursors, moisture, and other contaminations. In the typical procedure, we have chosen molybdenum trioxide ($MoO_3$) to sulfur (S) particle ratio ∼1:30 ($MoO_3$ = 15 mg, S = 100 mg). For the growth of $MoS_2$, 100 mg of NaCl powder was mixed with $MoO_3$ powder in a quartz boat and placed in the middle of the CVD furnace tube. Another boat was placed 13.5 cm away from the middle of the tube, which contains S powder. The $MoS_2$ was successfully grown at 650 °C for 20 min. Large-area growth of 3L-$MoS_2$ was achieved by controlling the concentration boundary layer formation in NaCl-assisted CVD of $MoS_2$. The concentration boundary layer is composed of sulfur and $MoO_3$ reactants, which are reacting in a gaseous phase under the influence of NaCl powder. However, to investigate the role of the concentration boundary layer on CVD growth, we have tuned the distance between $MoO_3$+NaCl precursors and the growing substrate. We have synthesized high-quality large-area (centimeter scale) 3L-$MoS_2$ film over the $SiO_2$/Si substrate with good repeatability in synthesis. As reported in our previous work, a water-soluble layer ($Na_2S$/$Na_2SO_4$) was also grown underneath $MoS_2$ [29]. $Na_2S$/$Na_2SO_4$ layer function as seed promoter and support the nucleation of large-area, uniform, and continuous $MoS_2$ film.

**2.2. Synthesis of Large-area Monolayer $WS_2$**

The same single-zone APCVD system was used to grow the large-area monolayer (1L) $WS_2$ over the sapphire substrate. The tube was purged at 150 °C with 480 sccm Ar gas flow for 30 minutes to remove humidity and pre-deposited contaminants. A mixture of NaCl and $WO_3$ (99.995 %, Sigma-Aldrich, 204781) powder was placed in the middle of the CVD furnace for 1 minute at 820 °C and 200 mg Sulfur (99.98%, Sigma-Aldrich, 414980) was placed at 20 cm away from the center of the furnace to achieve sulfurization in 120 sccm Ar gas flow environment.

### 2.3. Characterization Techniques

We have performed photoluminescence (PL) and Raman measurements at room temperature (RT) using Horiba Scientific (LabRAM HR Evolution) with 514 nm laser wavelength to study the optical properties of 3L $MoS_2$ and 1L $WS_2$ films. To analysis the surface morphology and thickness of the film, atomic force microscopy (AFM) of Bruker (Dimension ICON) was used. A monochromatic Al Kα X-ray line (probe size ~ 1.7 mm × 2.7 mm energy 1486.7 eV) was used for X-ray photoelectron spectroscopy (XPS) analysis. Philips Xpert Pro system with Cu Kα (λ= 1.54 Å) was used to perform the X-ray diffraction measurements (XRD). FESEM-Zeiss microscope (backscattering mode) was used to perform field emission scanning electron microscopy (FESEM). The photocurrent measurements were performed using a DC probe station (EverBeingEB6) coupled with a semiconductor characterization system of Keithley (SCS4200). A Xenon lamp (75 W) was used to measure the photoresponse of the device, which is combined with a computer interfaced monochromator (Bentham TMC-300V). Thorlabs power meter (PM-100D) was used for the power spectrum of the Xenon lamp.

### 2.4. Water-soluble Transfer Process

The complete layer transfer process is schematically illustrated in **Figure 1**. The substrate with as-grown 2D TMDC was first spin-coated by PMMA for 120 s at a speed of 1000 rpm. The assembly was kept at room temperature (RT) overnight for better adhesion of PMMA and TMDC. Afterward, complete assembly was dipped into the solution of 0.5 M NaOH. Before dipping into the solution, one edge of PMMA/TMDC was scratched so that NaOH solution can easily penetrate from there to the interface of growth substrate (GS) and TMDC. Within one minute of dipping in the solution, the PMMA/TMDC stack started to lift off from the GS and floating onto the surface of the solution. The dissolution of the Na$_2$S/Na$_2$SO$_4$ and the hydrophobic nature of TMDC begin the lift-off process. The PMMA/TMDC stack was then rinsed in DI water to remove the contaminations from the NaOH solution and transferred onto the target substrate (TS). After that, the PMMA/TMDC/TS assembly was blown by N$_2$ gas to evaporate the water molecules and again kept overnight so that transferred TMDC film gets better adhesion with TS. The hot acetone removed the PMMA, but still, there were some PMMA residues over the TMDC film. To remove these residues, the transferred film was placed in the Ar flow of 480 sccm at 350 °C for 2 h. Also, in this transfer process, we have selectively chosen 0.5 M NaOH solution for lift-off instead of 2M NaOH and pure hot DI water. In 2M NaOH solution, we observed that transferred film was highly damaged due to the high corrosivity of etchant (Figure S1). Similarly, when the complete PMMA/TMDC/TS stack was treated with hot DI water for lift-off, cracks and wrinkles were generated in the transferred film, worsening the film quality (Figure S2). Also, delamination of PMMA/MoS$_2$ stack occurs in more than 15 minutes in water.

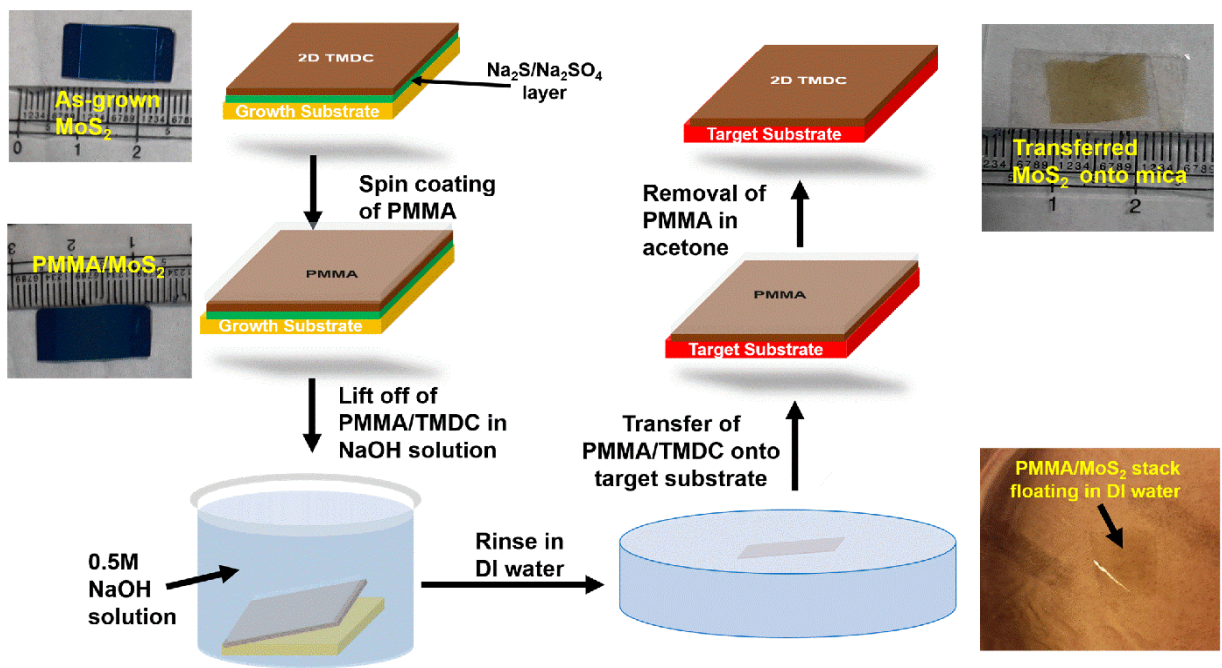

**Figure 1.** Schematic illustration of the water-soluble layer transfer process developed for MoS$_2$ and WS$_2$.

## 3. Result and Discussion

We can readily transfer large-area TMDCs using the aforementioned transfer process onto arbitrary substrates. To illustrate the viability of this process, we transferred centimeter-scale CVD-grown trilayer (3L) $MoS_2$ onto $SiO_2$/Si and sapphire substrates. **Figure 2(a)** shows the transferred 3L $MoS_2$ film from $SiO_2$/Si to the sapphire substrate, which indicates the complete lift-off and release of the film from the GS. Through the optical microscope (OM), we observed that transferred films are clean, continuous, and uniform with no cracks and wrinkles (**Figure 2(b, c)**). SEM and AFM images further confirm the clean nature and uniformity of the transferred $MoS_2$ film. SEM image clearly shows that the transferred 3L $MoS_2$ film is clean and wrinkle-free without polymer residues over it (**Figure 2(d)**). From the AFM analysis, it was observed that the surface roughness of transferred 3L $MoS_2$ film was comparable to the as-grown film. The film thickness was measured around 1.9 nm from the line scan along the white line, confirming the trilayer nature of $MoS_2$ film (**Figure 2(e)**).

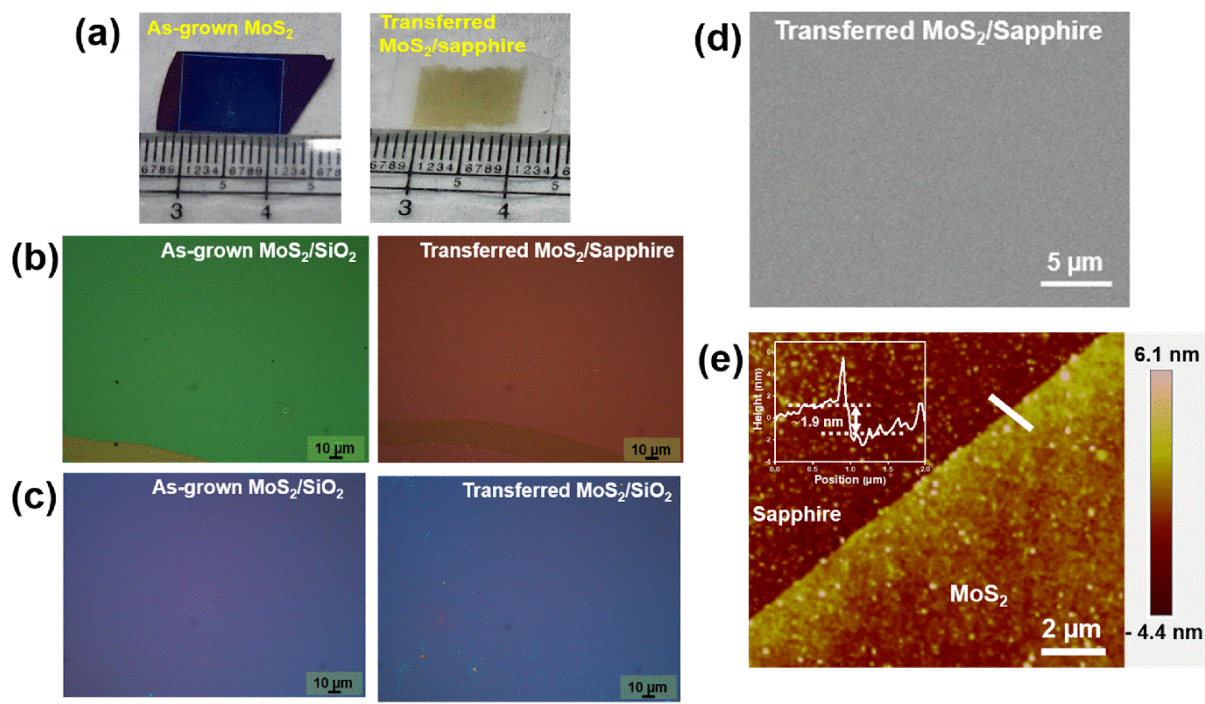

**Figure 2.** (a) Camera images of centimeter-scale as-grown and transferred 3L MoS$_2$. (b) Optical microscopy (OM) images of 3L MoS$_2$ synthesized on SiO$_2$/Si substrate and MoS$_2$ transferred onto the sapphire substrate. (c) OM images of as-synthesized 3L MoS$_2$/SiO$_2$ (GS) and transferred MoS$_2$/SiO$_2$ (TS). (d) SEM image of 3L MoS$_2$ transferred onto the sapphire substrate. (e) AFM image of transferred 3L MoS$_2$ onto the sapphire substrate. Inset: The thickness of 3L MoS$_2$ is shown by the height profile.

We examined the optical quality of the transferred 3L MoS$_2$ using Raman spectroscopy, which is widely used to investigate the structural and layer properties of 2D materials [30]. **Figure 3** depicts the Raman spectra of as-synthesized and transferred 3L MoS$_2$ film. Raman spectra of MoS$_2$ consist of two fundamental vibrational modes: A$_{1g}$ and E$^1_{2g}$, corresponding to out-of-plane and in-plane vibrations of atoms, respectively [22]. The peak position of these vibrational modes mainly depends on layer numbers and strain in the materials [31, 32]. The frequency difference ($\Delta\omega$) between A$_{1g}$ and E$^1_{2g}$ is an indicator of the number of layers in MoS$_2$. For as-grown MoS$_2$, we observed that $\Delta\omega$ is 22.3 cm$^{-1}$, indicating that the MoS$_2$ film is trilayer in nature [33]. It is noteworthy that $\Delta\omega$ changes slightly from 22.3 cm$^{-1}$ to 22.6 cm$^{-1}$ and 22.3 cm$^{-1}$ to 22.5 cm$^{-1}$ transferring 3L MoS$_2$ film onto SiO$_2$/Si (TS) and sapphire substrate, respectively (**Figure 3(a, b)**). This indicates that no strain was introduced in the film during the transfer process [34]. The small shift occurs due to the change in the interaction between film and substrate rather than the change in layer numbers and strain-induced during the transfer. The as-grown film strongly interacts with the substrate as compared to the transferred film [34]. From the Raman and AFM measurements, it has been confirmed that no MoS$_2$ is left on GS after the transfer, which ensures the complete lift-off of the MoS$_2$ from the GS (Figure S3, S4). From the AFM measurements, we have already confirmed that 3L MoS$_2$ remains as 3L after the transfer (**Figure 2(e)**).

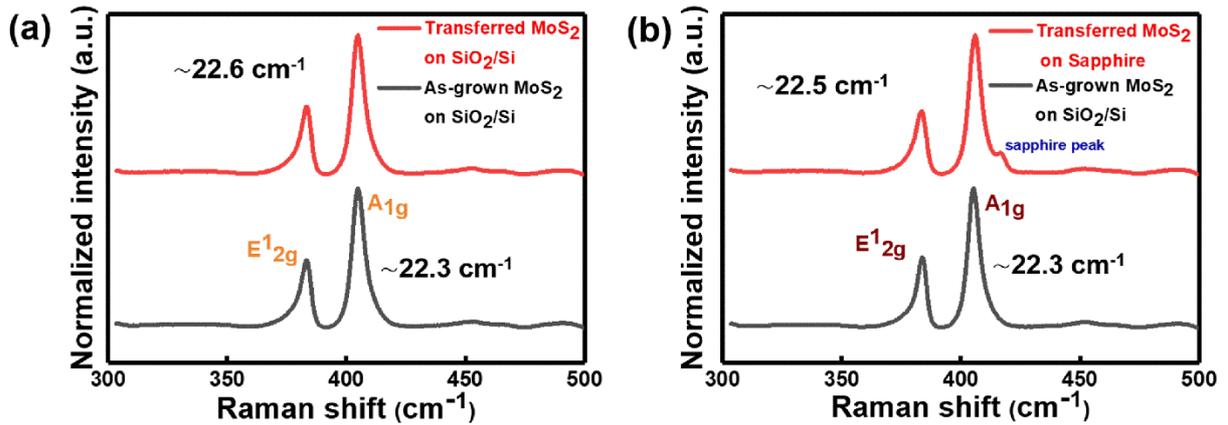

**Figure 3.** Raman spectra of 3L MoS$_2$ transferred from (a) SiO$_2$/Si (GS) to SiO$_2$/Si (TS) and (b) SiO$_2$/Si to sapphire substrate.

We did XPS analysis of the as-grown and transferred 3L MoS$_2$ (GS-SiO$_2$/Si; TS-SiO$_2$/Si) to understand the effect of transfer on the stoichiometry and chemical composition of MoS$_2$ film. **Figure 4(b)** shows the core-level spectra of Mo 3d and S 2s orbitals of the as-grown 3L MoS$_2$ film. The Mo$^{+4}$ 3d spectra consist of two peaks centered at 229.8 and 233.0 eV, corresponding to 3d$_{5/2}$ and 3d$_{3/2}$, respectively. The FWHM of 3d$_{5/2}$ and 3d$_{3/2}$ peaks are ~0.65 and ~0.87 eV, respectively. A small hump at 227.0 eV in the spectra, besides the peaks of Mo 3d, corresponds to S 2s. The spectra of S 2p deconvoluted into two peaks of 2p$_{3/2}$ and 2p$_{1/2}$, centered at 162.7 and 163.8 eV, with FWHM of ~0.65 and ~0.68 eV, respectively (**Figure 4(c)**). All the values are correctly matching with the previous literature reports [35, 36]. As shown in Figure **4(e, f)**, it has been observed that all the peak positions and FWHMs of the transferred film are almost similar to the as-grown film, indicating no change in the stoichiometry and chemical composition 3L MoS$_2$ film after the transfer. The presence of Na$_2$S/Na$_2$SO$_4$ underneath MoS$_2$ film was confirmed by Na 1s core-level spectra of as-grown film positioned at 1071.2 eV [29, 37] (**Figure 4(g)**). However, no Na 1s peak was observed in the transferred 3L MoS$_2$, confirming the dissolution of Na$_2$S/Na$_2$SO$_4$ layer during the transfer process (**Figure 4(h)**). **Figure 4(i)** shows the water contact angle of transferred 3L MoS$_2$/SiO$_2$.

The contact angle was around ~94.0°, which clearly shows the hydrophobic nature of MoS$_2$. The surface energy was calculated by the Fowkes model using the contact angles of diiodomethane and DI water [38]. The total surface energy of 3L MoS$_2$ was 35.72 mN/m, which is the sum of two components: polar component (0.72 mN/m) and dispersive component (35.00 mN/m). The dispersive component arises from the induced dipole-dipole interaction, and the polar component originates from the permanent dipole-dipole interaction [39].

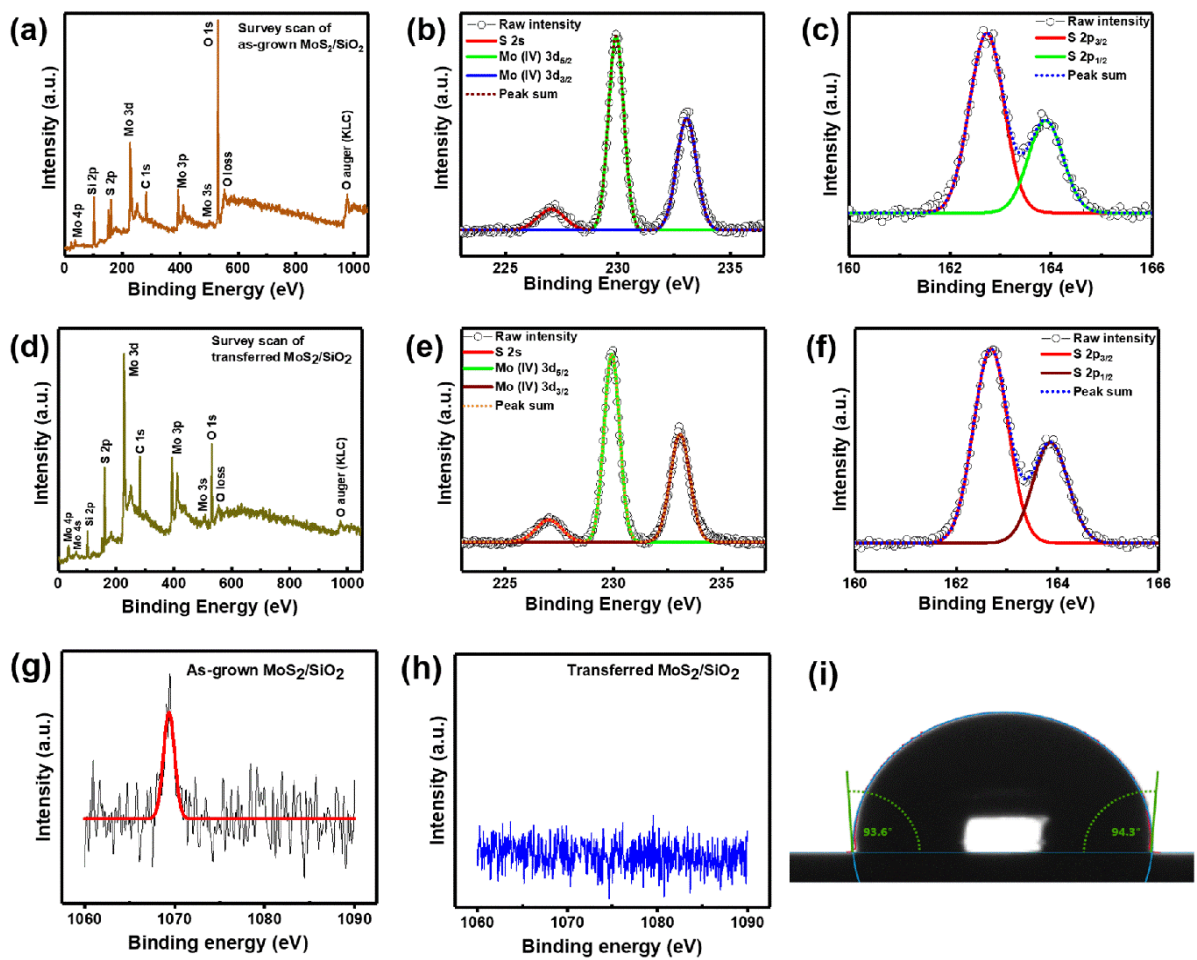

**Figure 4.** XPS spectra of 3L MoS$_2$ synthesized on SiO$_2$/Si (GS) and transferred onto SiO$_2$/Si (TS). (a, d) Survey spectrum of synthesized and transferred MoS$_2$, respectively. Core level spectra of (b, e) Mo 3d, and (c, f) S 2p peaks of as-grown and transferred MoS$_2$, respectively. (g) Na 1s binding energy spectrum of as-grown MoS$_2$, which signifies the presence of Na$_2$S/Na$_2$SO$_4$ layer underneath MoS$_2$ film. (h) Na 1s peak was not observed in the transferred

3L MoS$_2$, confirming the dissolution of Na$_2$S/Na$_2$SO$_4$ layer during the transfer process. (i) Contact angle measured for the transferred 3L MoS$_2$/SiO$_2$.

The optical properties and crystallinity of transferred 3L MoS$_2$ film were investigated by using the UV-Visible and XRD measurements on the transferred MoS$_2$/Sapphire. In the UV-Vis absorbance spectra of 3L MoS$_2$, we observed two prominent peaks at wavelengths 663 and 614 nm corresponding to A and B excitonic bands (**Figure 5(a)**). These two excitons get generated due to the direct transition between conduction band minimum and split valence band's maxima at the K point of the Brillouin zone [3, 23]. The absorbance peak (C exciton) around ~440 nm was also observed, which corresponds to van Hove singularities in the electronic density of states of MoS$_2$ [31, 40]. As shown in **Figure 5(b)**, the XRD analysis of 3L MoS$_2$ shows a broad peak corresponds to the (002) plane at Bragg's angle 2θ ≈ 13.7°. XRD pattern matches with the JCPDS card number 37-1492, confirming the hexagonal structure of MoS$_2$ with an interlayer spacing of 1.1 nm [41]. The selected area electron diffraction (SAED) pattern also ensures the single-crystalline nature and hexagonal symmetry of 3L MoS$_2$ film (**Figure 5(c)**). Therefore, UV-Visible and XRD confirm the high optical quality, uniformity, and single-crystalline nature of large-area 3L MoS$_2$ film.

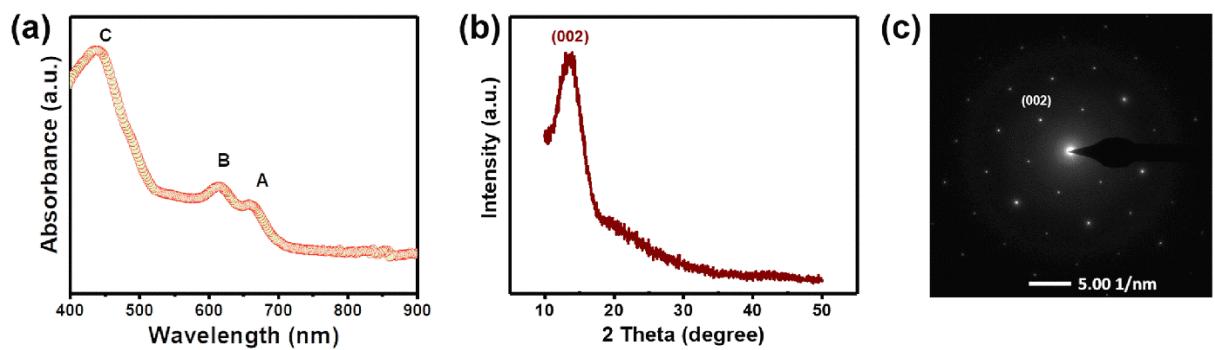

**Figure 5.** (a) UV-Visible absorbance spectra, and (b) XRD graph of transferred 3L MoS$_2$ onto the sapphire substrate. (c) SAED pattern showing the single-crystalline nature of 3L MoS$_2$.

### 3.1. Transfer onto flexible substrates

To enhance the potential application of TMDCs in flexible electronics, we have transferred 3L MoS$_2$ onto flexible substrates: Muscovite mica (highly flexible and high-temperature stable) and polyimide. Optical microscopic images clearly show the clean and wrinkle-free transferred films (**Figure 6(c, d)**). From the Raman measurements shown in **Figure 6(e)**, we found that Δω changes from 22.2 cm$^{-1}$ to 22.6 cm$^{-1}$ and 22.2 cm$^{-1}$ to 22.5 cm$^{-1}$ for MoS$_2$ transferred onto mica and polyimide substrate, respectively. This again indicates that no strain was introduced in the film during the transfer process [34]. We also did the XPS analysis of transferred 3L MoS$_2$/Mica to confirm that our transfer process does not change the stoichiometry and chemical composition of MoS$_2$ film. From **Figure 7(a, b)**, we observed no change in the peak positions and FWHM of peaks 3d$_{5/2}$, 3d$_{3/2}$, 2p$_{3/2}$, and 2p$_{1/2}$, which clearly shows that stoichiometry and chemical composition of transferred MoS$_2$ film consistent with the as-grown film. The significance of Na 1s peak was also not observed in the transferred MoS$_2$/Mica sample.

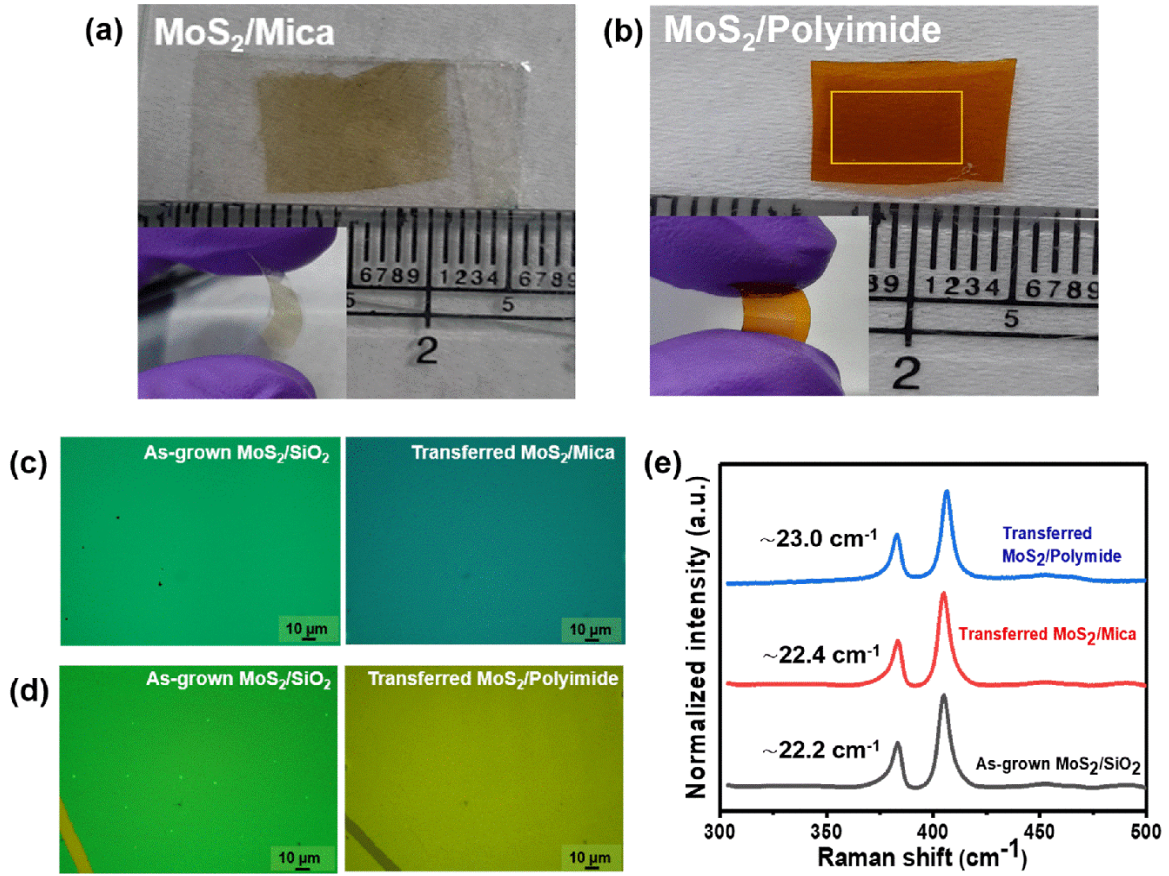

**Figure 6.** Photographs of large-area (centimeter scale) transferred 3L MoS$_2$ film onto flexible (a) mica and (b) polyimide substrate. The inset of (a) and (b) show MoS2/Mica and MoS2/Polyimide bending, respectively. (c) OM images of as-synthesized 3L MoS$_2$/SiO$_2$ and transferred MoS$_2$/Mica. (d) OM images of as-synthesized 3L MoS$_2$/SiO$_2$ and transferred MoS$_2$/Polyimide. (e) Raman spectra of as-grown and transferred 3L MoS$_2$ onto mica and polyimide substrates, respectively.

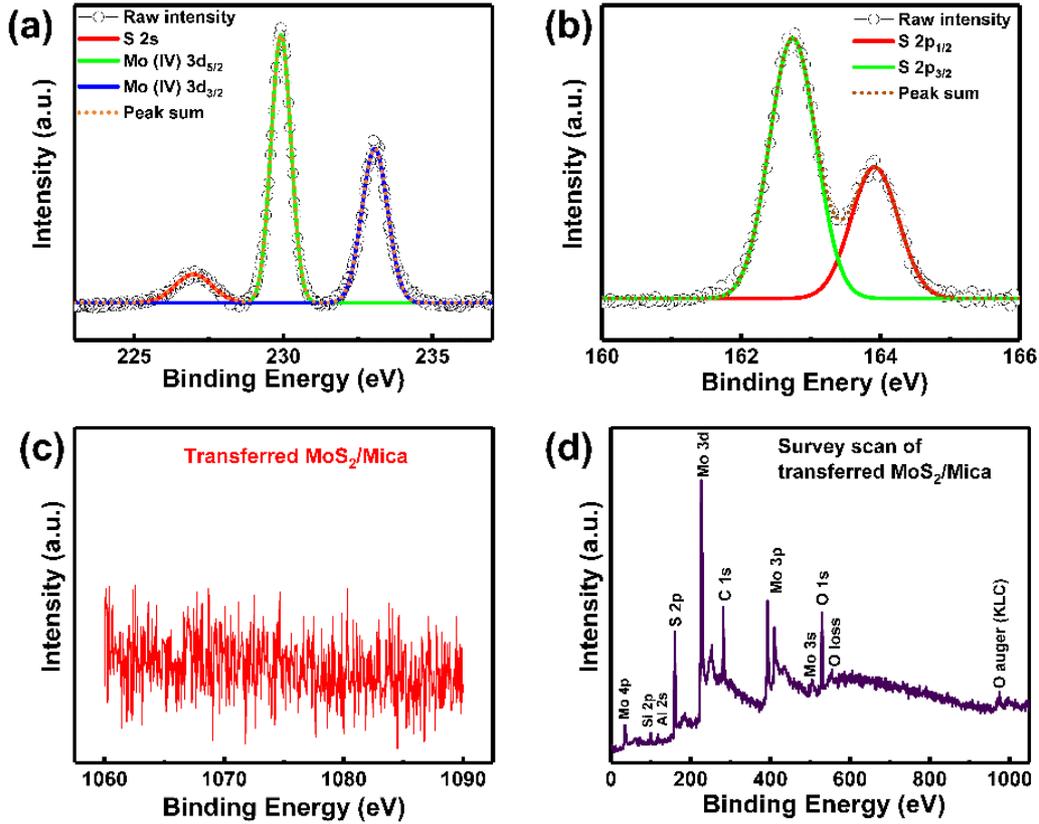

**Figure 7.** XPS spectra of core levels of (a) Mo 3d and (b) S 2p binding energies of transferred 3L MoS$_2$ onto mica substrate. (c) No significance of Na 1s peak in the transferred 3L MoS$_2$ (d) Survey scan of transferred 3L MoS$_2$/Mica.

### 3.2. Layer transfer of large-area monolayer WS$_2$

To show the universality and feasibility of the transfer process, we implemented our transfer process to the transfer of large-area monolayer WS$_2$ from sapphire to mica substrate. **Figure 8(a)** shows an as-grown sample of 1L WS$_2$ on the sapphire substrate where W1 and W2 are the monolayer regions. The monolayer WS$_2$ film has been transferred from W1, and W2 to WT1, and WT2, respectively (WT1 and WT2 are mica substrates). The OM images shown in **Figure 8(b)** are confirming the cleanliness of the transferred film. Raman and PL measurements were performed to examine the quality of transferred 1L WS$_2$. The Raman spectra of as-grown and transferred WS$_2$ are depicted in **Figure 8(c)**. Raman spectra contain two first-order vibrational modes: E$^1_{2g}$ (in-plane) and A$_{1g}$ (out-of-plane) mode. We also observed two peaks at 324.1 and 352.5 cm$^{-1}$ correspondings to 2LA(M)- E$^2_{2g}$ and 2LA(M) modes, respectively. The E$^1_{2g}$ mode of transferred WS$_2$ exhibits a blue shift of ~1.5 cm$^{-1}$ as compared to as-grown WS$_2$ due to the tensile strain release effect [42]. However, no shift was observed in A$_{1g}$ mode after the transfer because it is not impacted by strain. Conversely, A$_{1g}$ mode is susceptible to charge doping effect, while E$^1_{2g}$ mode is unaffected by the charge doping because of the strong electron-phonon coupling [43-45]. In this transfer process, no additional charge doping was introduced; as a result, the A$_{1g}$ mode of transferred and as-grown WS$_2$ remains identical. The 1L WS$_2$ shows the optical response from A and B excitonic transitions, which arise from the splitting of valence band maxima due to the spin-orbit interaction [46, 47]. In the PL spectrum of as-grown 1L WS$_2$, a single peak was observed at 1.99 eV, which may contain both charged (X$^-$ and X$^+$ trions) and neutral (X) A excitons (**Figure 8(d)**). The B excitonic peak is not measurable with this laser excitation. The PL spectrum of transferred WS$_2$ clearly shows two distinct components: neutral exciton (X) at 2.02 eV and negatively charged trion (X$^-$) at 1.99 eV, which are well fitted by the Lorentzian function. Note that the neutral exciton is fully absent in as-grown WS$_2$, while the X$^-$ trion peak of transferred WS$_2$ is very similar to as-grown WS$_2$ [44]. The suppression of neutral excitonic peak in as-grown WS$_2$ is due to the unintentional doping of electrons from the water-soluble layer underneath WS$_2$ and the n-type semiconducting nature of WS$_2$. The water-soluble layer was completely dissolved during the transfer, reducing the unintended electron doping in WS$_2$ [27]. This is the reason that the intensity of X exciton dominates the intensity of X- trion in the PL spectrum of transferred WS$_2$. Raman and PL spectra confirm that the 1L WS$_2$ retains its optical properties after the transfer.

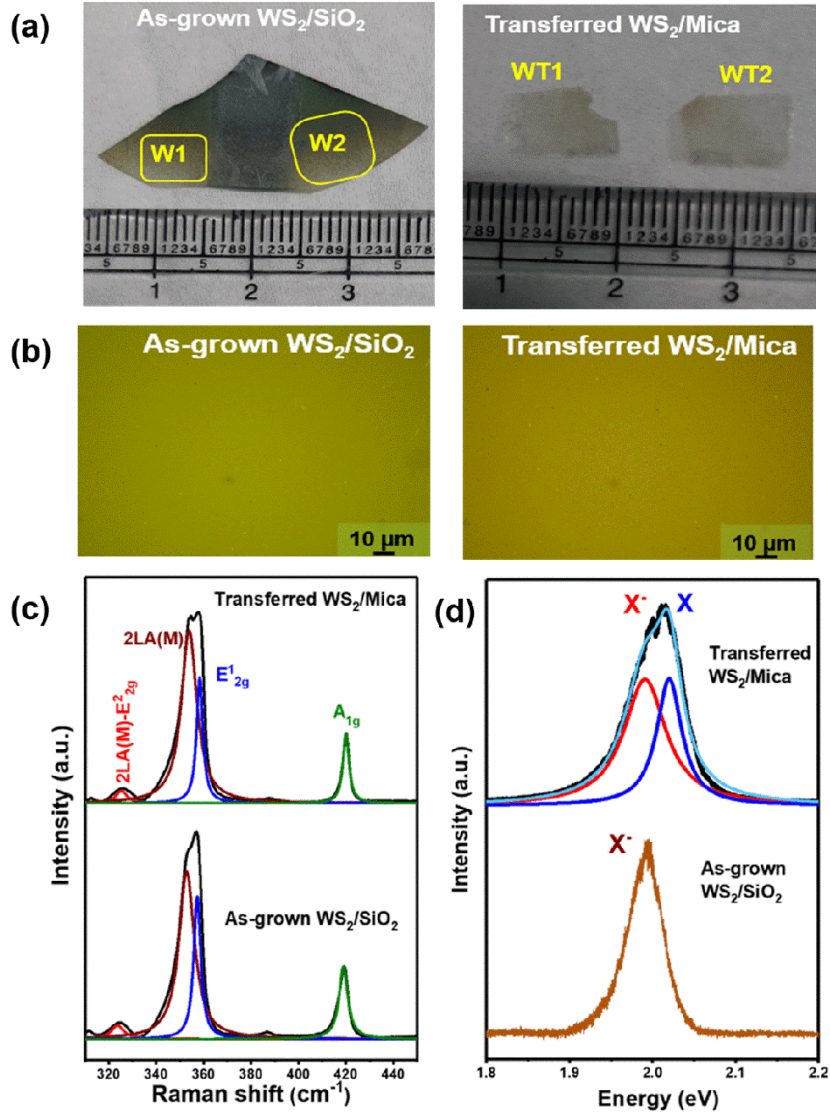

**Figure 8.** (a) Photographs of as-grown as well as transferred 1L WS$_2$ from sapphire to mica substrate. W1 and W2 are the monolayer regions of as-grown WS$_2$/Sapphire. Similarly, WT1 and WT2 are the mica substrates onto monolayer WS$_2$ was transferred from W1 and W2, respectively. (b) OM images of synthesized 1L WS$_2$/Sapphire and transferred 1L WS$_2$/Mica. (c) Raman and (d) PL measurement of as-grown and transferred WS$_2$.

## 3.3. Device application of Water-soluble transfer method

The compatibility of this transfer method with the nanoscale device fabrication technology was demonstrated by fabricating a broadband photodetector onto the transferred 3L

MoS$_2$/Sapphire. The typical schematic of metal-semiconductor-metal (MSM) photodetector is shown in **Figure 9(a)**. The electrical contacts of Ag/Au (40/60 nm) were deposited onto MoS$_2$ film by thermal evaporation using a metal mask. The photo-to-dark current ratio (PDCR), Responsivity (R), and Response time (t$_{res}$) are the key parameters for the performance of a photodetector. The photoresponse (*K*), Responsivity (R) are defined as

$$\text{PDCR} = \frac{I_p}{I_d} \quad \text{and} \quad R = \frac{I_p - I_d}{P_\lambda A_{eff}}$$

Where $I_d$ and $I_p$ are dark current and photocurrent, respectively. The power density is symbolized by $P_\lambda$ corresponding to wavelength $\lambda$. $A_{eff}$ is the effective area of the device illuminated with light, which is found to be 4.9 mm$^2$ for the present case. **Figure 9(b)** shows the room temperature dark and photocurrent at wavelength 250 and 650 nm with power density 1 and 23.34 µW/mm$^2$, respectively. A significant increment can be seen in the photocurrent upon illumination of light. Interestingly, the photocurrent increased 10 times to dark current for wavelength 650 nm (visible region) while it increased 100 times for wavelength 250 nm (UV region). The enhancement in the current can be attributed to the generation of electron-hole pairs when the device is exposed to light under a biasing voltage. **Figure 9(c)** shows the spectral responsivity measurements carried out from 230 to 800 nm wavelength at a fixed applied voltage (1.5 V). Responsivity continuously increases with a decrease in illuminated wavelength. At wavelength 800 nm (NIR region) corresponds to the bandgap ~1.5 eV of trilayer MoS$_2$, the device shows weak photoresponse because the power density used here is very small (12.65 µW/mm$^2$). As reported by *Woong Choi et al.*, high power density is required for the noticeable photoresponse at wavelength 800 nm due to the weak absorption tail of the indirect bandgap transition [48]. The first significant increment in responsivity was observed at 650 nm, corresponding to the bandgap of monolayer MoS$_2$. However, responsivity is significantly increased in UV region compared to visible reason, and the maximum value of R is 8.6 mA/W even at low applied voltage (1.5 V) and low power density (0.6 μW/mm$^2$). The obtained value of responsivity at relatively low applied voltage is higher than some of the previously reported UV photodetectors based on 2D materials [48-50]. The temporal response was taken to estimate the response speed of our device (**Figure 9(d)**). The fall time of current is relatively slow to the rise time, which may be due to the presence of traps and vacancies in the material. Present work paves the way to utilize the 2D TMDCs for nanodevice fabrication.

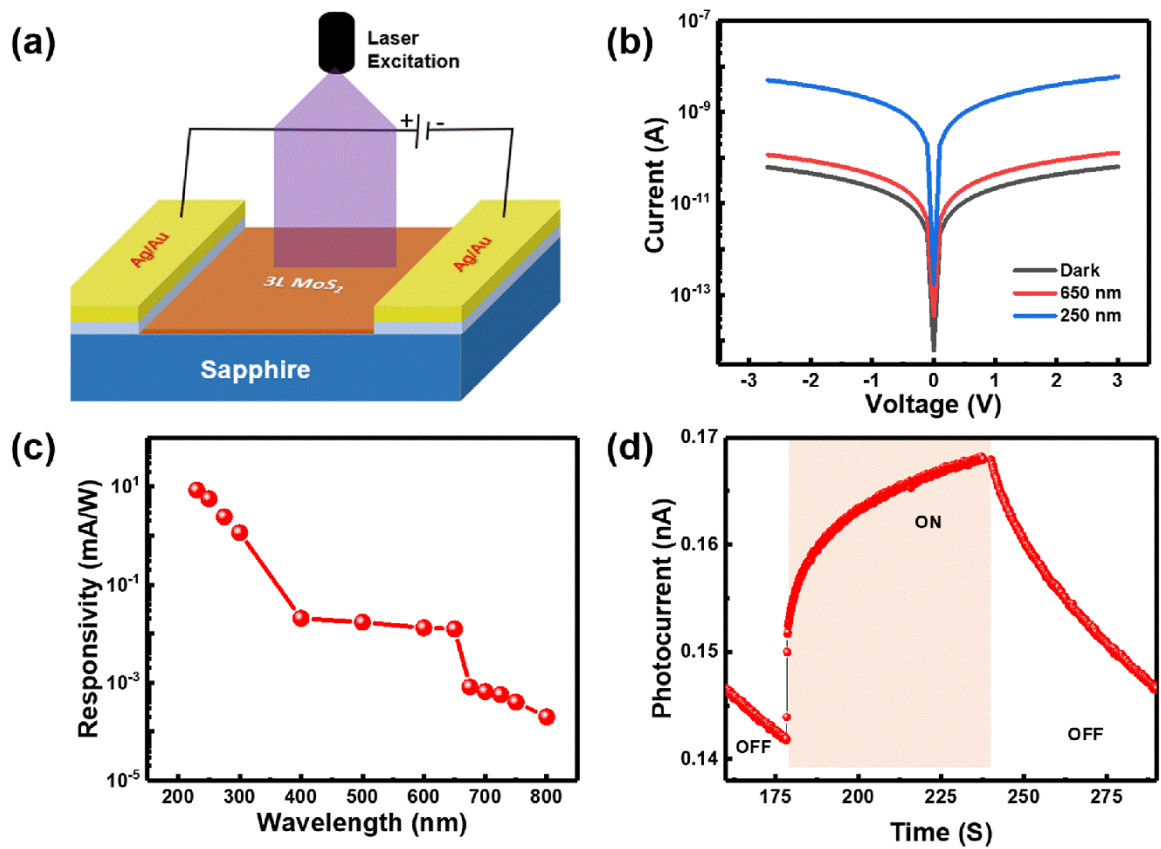

**Figure 9.** (a) Schematic illustration of MSM photodetector with a monochromatic light beam. (b) I-V characteristics of the photodetector in the dark and under 650 and 250 nm illuminating wavelengths. (c) The variation of responsivity of the device with illuminating wavelength ranging from UV to NIR. (d) Temporal response of the device recorded at 5 V biased voltage. The device was exposed to light for a period of 100 s, and then the light was turned off for 100 s.

## 4. Conclusion

In summary, we have demonstrated a water-soluble layer-based transfer method enabling the clean transfer of large-area (centimeter scale) CVD-grown 2D TMDCs onto arbitrary substrates, including flexible substrates such as mica and polyimide. The preservation of the crystalline quality of the transferred film was confirmed by performing various characterization techniques. The versatility of the transfer method has been shown by transferring 3L $MoS_2$ and 1L $WS_2$ onto different substrates. Large-area transfer of TMDCs onto flexible substrates (mica, polyimide) allows the fabrication of flexible devices. This water-soluble layer-based transfer technique can be a good alternative to the wet-etching transfer method. The photodetector fabricated onto transferred 3L $MoS_2$ shows the compatibility of this transfer method with the nanoscale device fabrication technology. In addition, the photodetector exhibits broadband photoresponse (NIR-Vis-UV) with a maximum responsivity of 8.6 mA/W. This work will serve the interest of the research community working towards the manufacturing of devices based on 2D materials for electronics, optoelectronics, bio-inspired electronics, and flexible electronics. This transfer process can also integrate 2D materials with various platforms such as Si augmentation/replacement, IoT, 2D/2D heterostructures.

## Acknowledgments

Madan Sharma thanks the Department of Science & Technology (DST) for the award of the research fellowship. The authors acknowledge the Nanoscale Research Facility (NRF) and Central Research Facility (CRF), Indian Institute of Technology Delhi, New Delhi, for providing the characterization facilities. We would also like to acknowledge 'Grand Challenge Project on MBE growth of 2D materials' sponsored by Ministry of Human Resource Development (MHRD), India, and IIT Delhi for partial financial support for this work.

**Data availability statement**

All data that support the findings of this study are included within the article (and in supporting files).

# Supporting information

# Large-Area Transfer of 2D TMDCs Assisted by Water-soluble layer for Potential Device Applications

*Madan Sharma[1], Aditya Singh[1], Pallavi Aggarwal[1] and Rajendra Singh[1,2*]*

[1] Department of Physics, Indian Institute of Technology Delhi, Hauz Khas, New Delhi 110016, India

[2] Nanoscale Research Facility (NRF), Indian Institute of Technology Delhi, Hauz Khas, New Delhi 110016, India

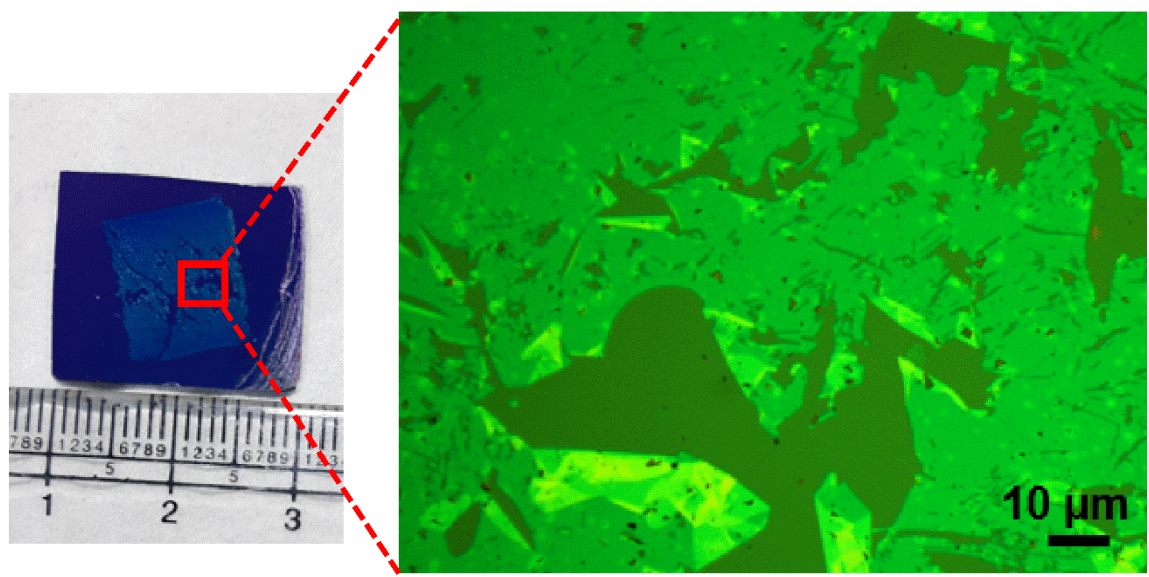

**Figure S1**. Photograph of transferred 3L MoS$_2$ lifted off in 2M NaOH solution. The optical image clearly shows that film was damaged during the transfer process.

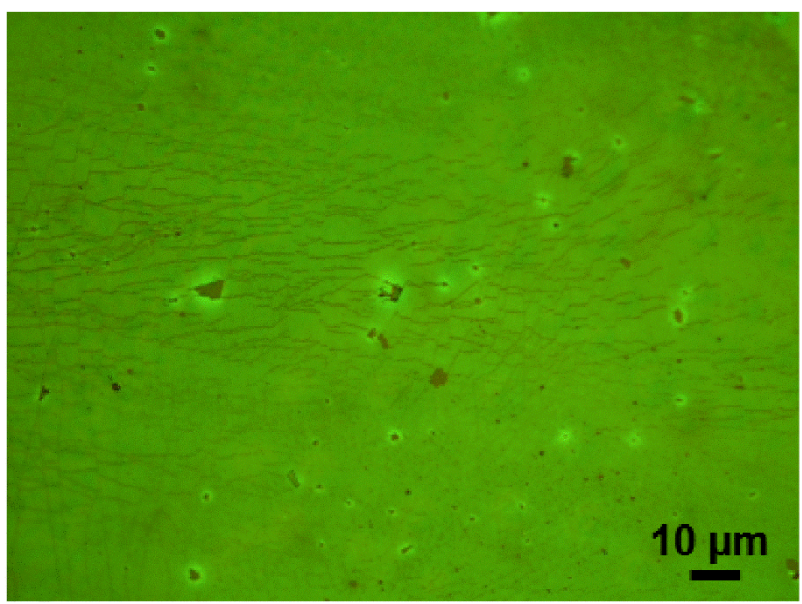

**Figure S2.** Optical image of transferred 3L MoS$_2$ lifted off in hot DI water. Cracks and wrinkles are formed in the transferred film due to the generation of water bubbles during the heating of water > 80 ℃. Also, PMMA/MoS$_2$ stack was delaminated in more than 15 minutes.

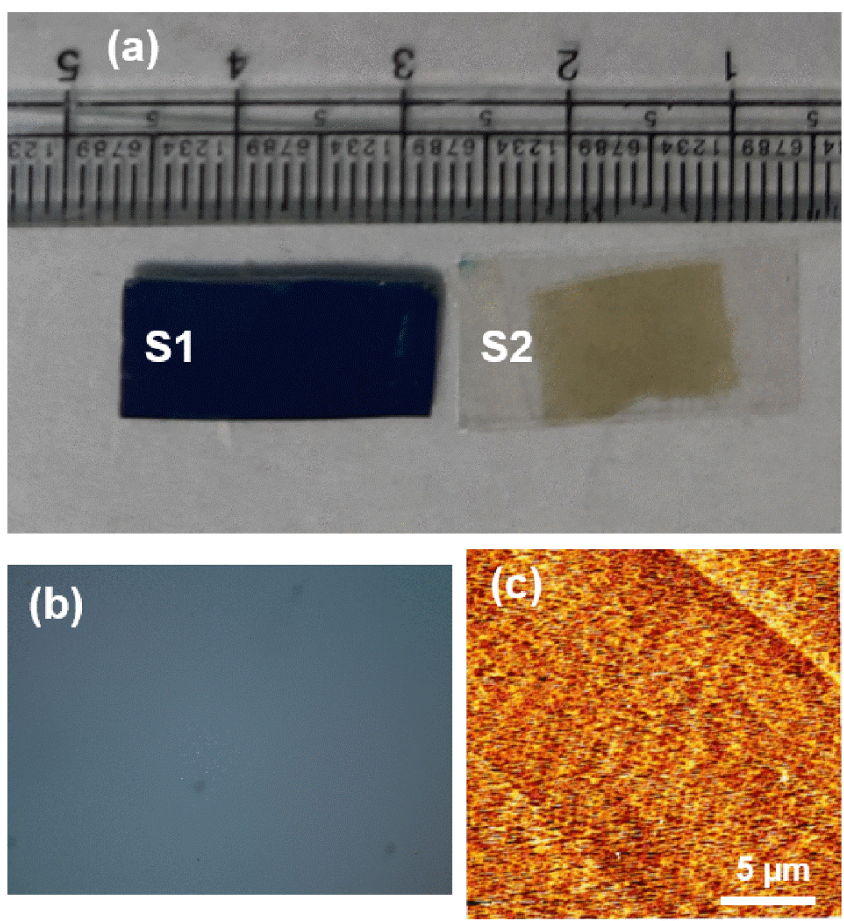

**Figure S3.** (a) Photograph of SiO$_2$/Si growth substrate (S1) from where trilayer MoS$_2$ has been transferred onto target substrate (S2). (b) and (c) are optical and AFM images of S1, respectively.

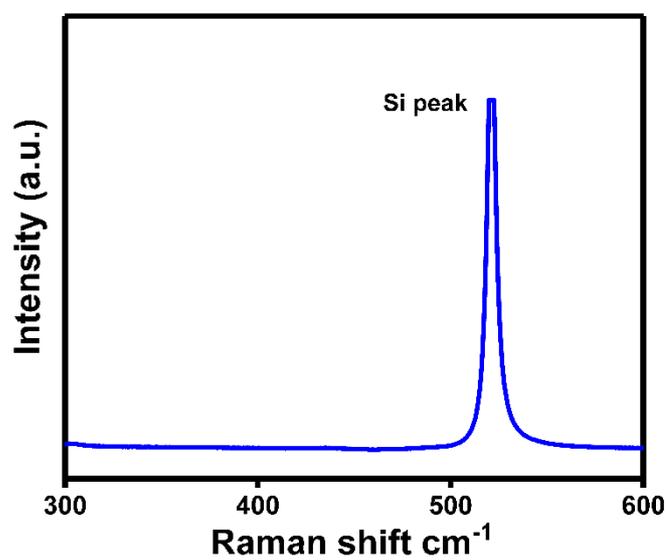

**Figure S4.** Raman spectra of sample S1. No $A_{1g}$ and $E^1_{2g}$ peaks were observed between 300 to 500 cm$^{-1}$, which confirms the complete lift-off of MoS$_2$ film from the growth substrate.